# Hydrogen-based direct reduction of multicomponent oxides: Insights from powder and pre-sintered precursors toward sustainable alloy design

Shiv Shankar[1], Barak Ratzker[1,*], Yan Ma[1,2,*], Dierk Raabe[1,*]

[1] Max Planck Institute for Sustainable Materials, Max-Planck-Str. 1, 40237, Düsseldorf, Germany

[2] Department of Materials Science & Engineering, Delft University of Technology, Mekelweg 2, 2628 CD Delft, the Netherlands

*Corresponding author: y.m.ma@tudelft.nl (Y. Ma); b.ratzker@mpi-susmat.de (B. Ratzker); d.raabe@mpi-susmat.de (D. Raabe)

## Abstract

The co-reduction of metal oxide mixtures using hydrogen as a reductant in conjunction with compaction and sintering of the evolving metallic blends offers a promising alternative toward sustainable alloy production through a single, integrated, and synergistic process. Herein, we provide fundamental insights into hydrogen-based direct reduction (HyDR) of distinct oxide precursors that differ by phase composition and morphology. Specifically, we investigate the co-reduction of multicomponent metal oxides targeting a 25Co-25Fe-25Mn-25Ni (at.%) alloy, by using either a compacted powder (mechanically mixed oxides) comprising $Co_3O_4$-$Fe_2O_3$-$Mn_2O_3$-NiO or a pre-sintered compound (chemically mixed oxides) comprising a Co,Ni-rich halite and a Fe,Mn-rich spinel. Thermogravimetric analysis (TGA) at a heating rate of 10 °C/min reveals that the reduction onset temperature for the compacted powder was ~175 °C, whereas it was significantly delayed to ~525 °C for the pre-sintered sample. Nevertheless, both sample types attained a similar reduction degree (~80%) after isothermal holding for 1 h at 700 °C. Phase analysis and microstructural characterization of reduced samples confirmed the presence of metallic Co, Fe, and Ni alongside MnO. A minor fraction of Fe remains unreduced, stabilized in the (Fe,Mn)O halite phase, in accord with thermodynamic calculations. Furthermore, ~1 wt.% of BCC phase was found only in the reduced pre-sintered sample, owing to the different reduction pathways. The kinetics and thermodynamics effects were decoupled by performing HyDR experiments on pulverized pre-sintered samples. These findings demonstrate that initial precursor states influence both the reduction behavior and the microstructural evolution, providing critical insights for the sustainable production of multicomponent alloys.

## 1 Introduction

The production of metals and alloys accounts for approximately 40% of all industrial greenhouse gas emissions, mainly due to the reliance on fossil-based reductants [1,2]. Therefore, decarbonization of the metallurgical sector offers a huge leverage for pursuing global carbon neutrality targets [3]. Hydrogen-based direct reduction (HyDR) of metal oxides has emerged as a promising pathway for sustainable metal production, as it uses green hydrogen as reductant and produces water as the redox product [4–10].

The co-reduction of metal oxide mixtures using hydrogen offers an innovative route for sustainable alloy design with targeted microstructures in a single process [11–17]. However, the initial structural and chemical state of the oxide precursors, such as, (i) mechanically mixed oxides, (ii) chemically mixed oxides (pre-sintered), or (iii) natural ores, significantly influences the driving forces and kinetic solid-state reduction pathways [11,18,19]. With the former term we are referring here to a state where the individual oxides (here: $Co_3O_4$, $Fe_2O_3$, $Mn_2O_3$, and NiO) are only mechanically mixed, remaining chemically distinct and unmixed prior to the ensuing co-reduction. With the second term we are referring to a state where the oxides are chemically mixed at the atomic-scale, due to sintering in the inert atmosphere. The third term is important to distinguish from the other two (well-defined) oxide states, as natural ores are typically complex, chemically mixed mineral compounds (with the exception of most iron ores, which are concentrated with an Fe content up to ~67 wt.%). In many cases, such as lateritic Ni and Co weathering ores, the actual metal content of natural mineral feedstock is low, often not exceeding 1-1.5 wt.% in the targeted metals [20–22]. These differences in the initial oxide states not only influence the diffusion pathways but also the corresponding chemical driving forces under co-reduction conditions. This distinction is therefore critical when conducting reduction studies aimed at understanding the thermodynamics and kinetics of one-step alloy production, where reduction and alloying are integrated into a single process [17].

Mechanically mixed oxide powders are typically highly porous (20-30%), a state which promotes faster reduction kinetics due to better access to free surfaces for $MO_{(s)}+H_{2(g)} \rightarrow M_{(s)}+H_2O_{(g)}$ gaseous exchange [23]. In contrast, pre-sintered compounds and natural ores are denser and considerably less porous (5-15%), resulting in slower overall kinetics during HyDR processes [24]. As known from earlier hydrogen-based solid-state reduction studies, the progressive loss of oxygen during reduction leads to the formation of a relatively fine pore (and crack) network [25]. With continued reduction, this network evolves into a percolating internal free-surface structure, which in turn facilitates and accelerates the reduction process [26]. Depending on the duration of the high temperature treatment, the pore structure in the final compacted metallic alloy can be gradually healed out, due to capillary-driven sintering processes [27].

Beyond these profound differences in initial microstructures and the associated kinetics, recent experimental insights into HyDR of binary oxide mixtures have also highlighted the thermodynamic effects of the respective oxide precursor states on the overall reduction behavior. As an example, the hydrogen reduction of $NiO+Cr_2O_3$ exhibited major weight loss already below 300 °C [28], while the onset in the reduction of a $NiO+NiCr_2O_4$ solid solution occurred only above 300 °C. This change indicates that chemically mixed oxides can exhibit a changed reduction onset compared with mechanically mixed systems. Furthermore, the reduction sequence and mechanisms also differ depending on the precursor state [18]. In the mechanically mixed $Cr_2O_3+Fe_2O_3$ oxide mixture, reduction follows the thermodynamic stability of the individual oxides, as predicted by the Ellingham diagram [29]: $Fe_2O_3$ being less stable, reduces first to Fe, while $Cr_2O_3$ remains unreduced. Upon further heating to 1100 °C, $Cr_2O_3$ is reduced, yielding a Fe-Cr alloy. In contrast, chemically mixed Fe-Cr-O solid solution undergoes an initial reduction to $FeCr_2O_4$ and precipitates $Fe_3O_4$ (which ultimately reduces to metallic Fe). The remaining $FeCr_2O_4$ oxide is reduced to Fe and precipitates $Cr_2O_3$ to Cr, forming a Fe-Cr alloy. These observations underscore the critical role of the precursor state in governing both the kinetics, thermodynamics, reduction mechanisms as well as the transient and resulting microstructures. Despite the growing interest in HyDR, the decoupling of thermodynamics and kinetics during hydrogen reduction of oxide mixtures and the leverage of such studies on sustainable one-step alloy production remains a completely underexplored field to date. This gap is particularly significant for two reasons: first, to better understand the reduction behavior of naturally mixed mineral oxides commonly used in industry with the aim to explore if there are energetically (i.e. thermodynamically) favorable pathways for the direct production of multicomponent alloys by co-reduction. Second, we try with such insights to discover, design and advance one-step metallurgical production pathways,, in which stoichiometrically mixed and adequately pre-blended oxide precursors can be directly transformed into final alloys through a single integrated step of reduction, compaction, and alloying, the so-called one step metallurgy [30].

In this work, we investigate the HyDR behavior (100% $H_2$ at 700 °C) of (a) mechanically mixed, compacted powder samples and of (b) pre-sintered, chemically mixed variants of multicomponent oxides mixtures. More specifically, we target a 25Co-25Fe-25Mn-25Ni (at.%) multicomponent alloy system, comprising three easy-to-reduce metal oxides and the hard-to-reduce $Mn_2O_3$. The thermodynamics and kinetics effects influencing the HyDR process were decoupled by a multifaceted approach combining thermogravimetric analysis , phase analysis, microstructure characterization, and thermodynamic modeling. We find that the pre-sintered sample exhibits a delayed onset of reduction. This is attributed to both thermodynamic and kinetic effects resulting from preceding sintering process. Thermodynamically, formation of stable spinel and halite phases makes the reduction more difficult and kinetically, the absence of initial porosity hinders gas transport and slow down the reduction process.

## 2 Materials and experimental methods

### 2.1 Powder preparation

Four different metal oxides, namely, $Co_3O_4$, $Fe_2O_3$, $Mn_2O_3$, and NiO with ≥ 99% purity were weighed with a targeted equiatomic metallic concentration of 25 at.%. A powder mixture batch of approximately 14 g was weighed for mixing, as shown in **Table 1**. The powders were weighed using a balance with an accuracy of ±0.0001 g. The measured oxide powders were mixed and homogenized using a planetary ball mill (Fritsch 7) using hardened stainless-steel balls and crucibles. The ball milling was performed at 250 rpm for 15 cycles, with each cycle lasting for 20 min with a 5 min pause in-between cycles (5 h). Thereafter, the powder mixture was thoroughly removed from the crucible using ethanol solution, which was subsequently dried in an oven at 105 °C for 1 h.

Table 1. Composition of metal oxides in the prepared powder mixture.

| Metal oxide | Targeted metal content in alloy (at.%) | Metal in each oxide (wt.%) | Targeted metal content in alloy (wt.%) | Weight of oxide (g) | Wt.% of oxide |
|---|---|---|---|---|---|
| $Fe_2O_3$ | 25 | 70.00 | 24.45 | 3.49 | 25.45 |
| NiO | 25 | 78.59 | 25.70 | 3.26 | 23.81 |
| $Co_3O_4$ | 25 | 73.43 | 25.80 | 3.51 | 25.58 |
| $Mn_2O_3$ | 25 | 69.61 | 24.05 | 3.45 | 25.16 |
| **Total** | 100 | -- | -- | 13.71 | -- |

## 2.2 Cold compaction

To make individual samples, a powder mixture of ~1.5 g was poured into a pressing die (hardened steel with tungsten carbide core) with an inner diameter of 13 mm and compacted using a hydraulic press at a force of 30 kN (~225 MPa). After pressing, a disc-shaped compacted powder sample was obtained.

## 2.3 Sintering

A batch of compacted powder samples was then pre-sintered to produce a solid solution of chemically atomic-scale blended metal oxide mixtures. Sintering was performed under protective atmosphere in a sealed furnace with tungsten heating elements. The furnace was purged with Ar by pumping and flushing for three continuous cycles. Thereafter, the furnace was evacuated to $10^{-5}$ torr using diffusion pump. The sample was heated with a heating rate of 5 °C/min up to 1100 °C. Once a temperature of 600 °C was reached, Ar was introduced into the chamber. The samples were subjected to a dwell time of 5 h, followed by gradual cooling to room temperature in the furnace. After sintering the diameter of the pellet shrank by 11.35%. Moreover, to enable decoupling of thermodynamic and kinetic effects during HyDR, some pre-sintered reference samples were crushed with a hammer and subsequently ball-milled for 15 and 30 cycles, referred to as 5 and 10 h pulverized pre-sintered samples.

## 2.4 Thermogravimetric analysis

Non-isothermal reduction tests were carried out for the compacted powder, pre-sintered samples, and pulverized pre-sintered samples. **Figure 1** presents a schematic overview that summarizes the experimental procedure. The experiments were conducted using an in-house thermogravimetric analysis (TGA) configuration with an accuracy of 0.1 μg [31]. The mass of each sample was also measured by a weighing balance (±0.0001 g) before and after the reduction experiments. The sample was placed in a quartz crucible with a dimension of 16 mm inner diameter and 10 mm height. The reduction was carried out inside a 40-mm quartz tube with a gas inlet at bottom and outlet on top. The crucible containing the sample was suspended from one arm of the balance by quartz suspension rod, long enough to reach the center of the heating zone of an infrared-light furnace. The temperature profile was measured by a type-K thermocouple inserted into the center of the reference iron pellet placed directly below the crucible. Before heating the furnace, the reaction chamber was flushed with Ar gas at a flow rate of 10 L/h for 10 min. The Ar inlet was then closed, and hydrogen gas (with a purity of 99.999%) was introduced at a flow rate of 10 L/h. The experiment was initiated only after the balance stabilized. The samples were heated from 20 to 700 °C with a heating rate of 10 °C/min, and held isothermally for 1 h at 700 °C. The mass change was continuously recorded at 1 s intervals. After the completion of the reduction reaction, the furnace was switched off and the sample was rapidly cooled to room temperature. The hydrogen flow was maintained until the end of the cooling period to avoid any reoxidation of the reduced sample.

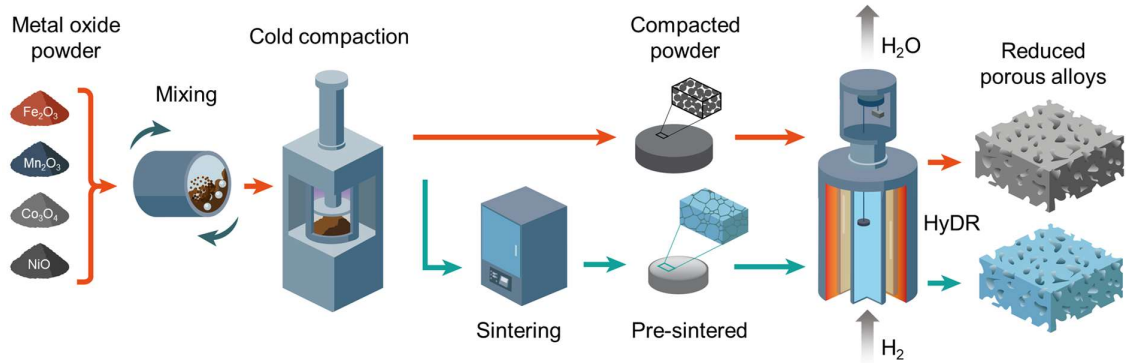

**Figure 1**. Schematic overview image depicting the compacted powder (mechanically mixed) and pre-sintered (chemically mixed) precursors subjected to hydrogen-based direction reduction (HyDR).

## 2.5    Materials characterization

The phase analysis of the initial and reduced samples was carried out by synchrotron high-energy X-ray diffraction (HEXRD). The measurements were performed at the beamline P02.1 Powder Diffraction and Total Scattering Beamline of PETRA III in Deutsches Elektronen-Synchrotron (DESY). The beamline was operated at an energy of 60 keV ($\lambda$=0.207381 Å). The sample was placed between the incident beam and the fast area detector Varex XRpad 4343CT (2880 pixels × 2880 pixels). The powder sample was glued onto a transparent tape and was positioned on a specimen stage, while the pre-sintered sample was placed directly on the specimen stage. The sample-to-detector distance and probing beam size was ~ 970 mm and 1×1 mm$^2$, respectively. A quarter of Debye–Scherrer diffraction rings (azimuthal angle: 180-270°) was recorded by a fast area detector. The total exposure time of each measurement was 5 s. The resulting diffraction ring patterns were integrated into 2D profiles using GSAS-II software [32], with a 2θ range of 1-18°. Moreover, the phase analysis of pulverized pre-sintered samples was carried out by X-ray diffraction using a Rigaku Smartlab 9kW diffractometer, equipped with a Cu K$_\alpha$ radiation ($\lambda$ = 1.5405 Å), operated at 45 kV and 200 mA. The 2θ scanning range was from 10° to 120° with a scanning step of 0.01°. The data post-processing and plotting were performed using Origin Pro 2022 software. The phases were determined according to the PDF-4+ database [33]. A quantitative phase analysis was performed by a pseudo-Voight function Rietveld refinement using the MDI JADE 10 software package.

The microstructures of all samples, before and after the reduction, were characterized by secondary electron (SE), backscattered electron (BSE), energy dispersive X-ray spectroscopy (EDS), and electron backscatter diffraction (EBSD) techniques using a ZEISS Sigma 500 scanning electron microscope (SEM). The samples were cut along the cross section of the pellets, embedded, and mechanically ground using silicon carbide papers from 400 to 2500 grits. The samples were then polished using diamond suspension (particle size 3 μm). Final polishing was done using 30 vol.% $H_2O_2$-contained colloidal silica suspension solution (particle size 50 nm). The acceleration voltage, and step size for EBSD measurements were 15 kV and 0.05 μm, respectively. The data were analyzed using OIM Analysis$^{TM}$ V9.0 software including spherical indexing [34].

## 2.6 Thermodynamic modeling

Thermodynamic calculations based on the CALPHAD approach were performed using Thermo-Calc software (version 2025a) to predict the thermodynamic stability in terms of the Gibbs free energy of metal oxide mixtures [35]. The thermodynamic modelling of the Co-Fe-Mn-Ni-O (multicomponent oxide) was performed using the TCOX10 metal oxide solutions database. The Gibbs free energy of the metal oxide mixture was calculated using equilibrium oxygen partial pressure between the metal oxide mixtures and the metallic phase [36]. The molar ratios of metals and oxygen were introduced in the equilibrium at a fixed temperature and gas pressure of 700 °C and 1 bar, respectively.

## 3 Results and discussion

### 3.1 Microstructure characterization of initial samples

The HEXRD patterns of the compacted powder and pre-sintered sample are shown in **Figure 2**. The diffraction peaks of the compacted powder sample correspond to individual metal oxides, namely $Co_3O_4$, $Fe_2O_3$, $Mn_2O_3$, and NiO. The quantified HEXRD phase fractions (**Table 2**) agree well with the targeted amount of the metal oxides in the initial oxide mixture (**Table 1**). This observation shows that no phase transformation and mechanical alloying took place during the preceding ball milling operation. In other words, the initial precursor was in the form of a compacted oxide agglomerate, devoid of any atomic-scale pre-mixing.

**Table 2**. Phase fractions (wt.%) of initial samples, measured by HEXRD with quantitative phase analysis using the Rietveld refinement.

| Sample | $Co_3O_4$ | $Fe_2O_3$ | $Mn_2O_3$ | NiO | Spinel | Halite |
|---|---|---|---|---|---|---|
| Compacted powder | 20.3±1.7 | 26.0±1.7 | 28.2±2.3 | 25.5±1.6 | -- | -- |
| Pre-sintered | -- | -- | -- | -- | 55.0±1.5 | 45.0±1.3 |

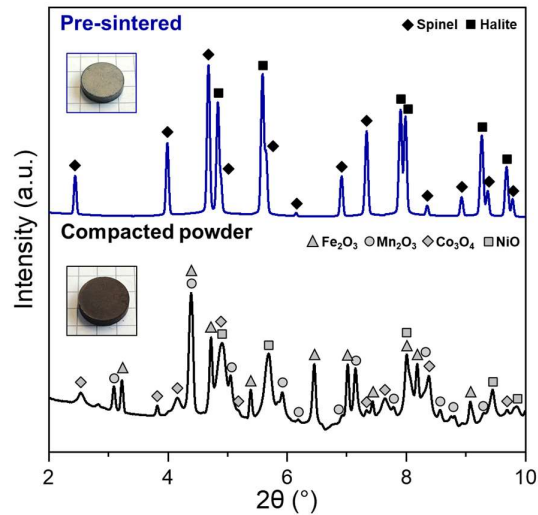

**Figure 2**. Synchrotron high-energy X-ray diffractograms (λ=0.207381 Å) of initial compacted powder and pre-sintered samples.

Microstructure of the compacted powder and pre- sintered samples are shown in **Figure 3**. The compacted powder consisted of loose particles and roughly ~26% porosity, **Figure 3a**. The corresponding EDS maps (**Figure 3c-f**) indicate a homogeneous distribution of Fe-, Co-, and Ni-oxide particles, while Mn oxide distributed non-uniformly with relatively larger particles. The particle size for Fe, Co, and Ni oxide was in a range of ~ 0.05-1 μm, whereas the majority of Mn oxide particles were ~ 0.2-6 μm.

Sintering of the mixed compacted powder sample resulted in the formation of a dual-phase microstructure comprising approximately 55 wt.% spinel and 45 wt.% halite phases (**Table 2**), as determined by HEXRD (**Figure 2**). The pre-sintered sample microstructure, consisting of halite and spinel phases, is presented in **Figure 3g**. The sample is ~98% dense with a small fraction of isolated pores found at triple points (see **Figure 3g** insert**)**, the larger faceted holes are grain pullouts that occur during metallographic preparation. Both phases (spinel and halite) had a similar grain size of about ~2 μm. The corresponding oxygen EDS map (**Figure 3h**) shows that the pre-sintered sample consists of oxygen-rich and oxygen-deficient phases, confirming the homogenous mixture of spinel ($M_3O_4$; 57.14 at.% O) and halite phases (MO; 50.00 at.% O). Further, EDS mapping revealed that Mn-Fe-rich regions correspond to the spinel phase (**Figure 3i** and **j**), while Co-Ni-rich regions (**Figure 3k** and **l**) to the halite phase.

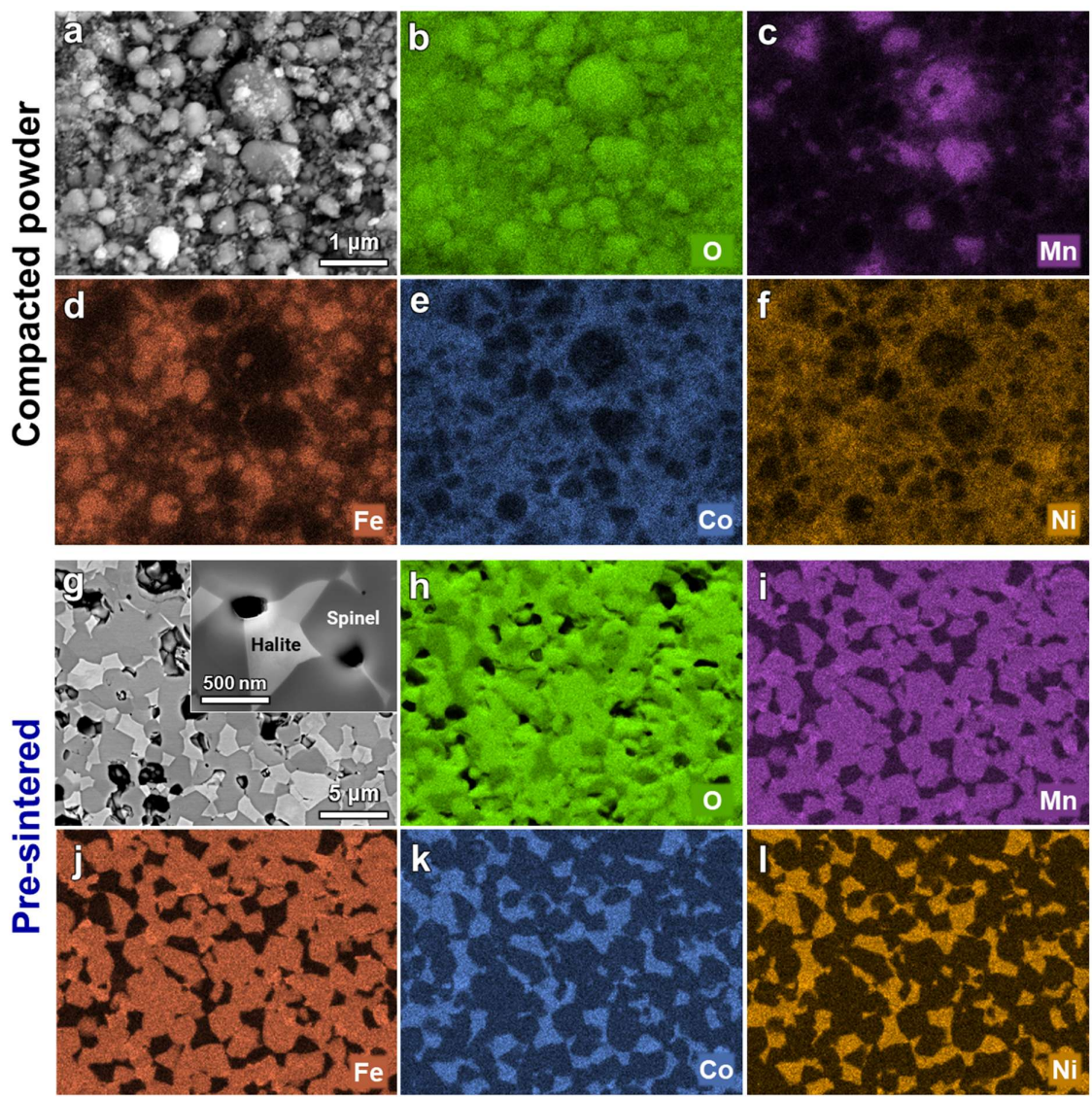

**Figure 3.** (a) SEM-SE micrograph of the compacted powder after the ball milling process and corresponding EDS elemental map of (b) O, (c) Mn, (d) Fe, (e) Co, and (f) Ni. (g) SEM-BSE micrograph of the pre-sintered samples (1100 °C in Ar atmosphere), the inset illustrates halite/spinel phases and isolated pores at higher magnification. Corresponding EDS mapping of (h) O, (i) Mn, (j) Fe, (k) Co, and (l) Ni, indicating the elements distribution in Fe-Mn-rich spinel and Co-Ni-rich halite phases.

## 3.2 Hydrogen-based direct reduction of compacted powder and pre-sintered sample

Prior to reduction, the compacted powder sample was subjected to a baking treatment by heating in an Ar atmosphere to a temperature of 300 °C, with a heating rate of 10 °C/min, and held isothermally at 300 °C for 30 min (**Figure S1**). The baking was performed to release moisture from the compacted powder, ensuring that the TGA mass loss recorded during the HyDR process only reflects oxygen removal.

**Figure 4a** shows the reduction degree of the compacted powder and pre-sintered samples as a function of time-temperature profile measured by TGA. The reduction onset temperatures of the two samples were remarkably different, that is, ~175 °C and ~525 °C for the compacted powder and the pre-sintered samples, respectively. In both cases there was a continuous mass loss during heating and essentially most of the reduction occurred before the isothermal holding stage. The compacted powder sample was nearly at its final reduction degree (0.76) at 620 °C, while the pre-sintered sample reached a much lower reduction degree (0.27) for the same temperature. The total reduction degrees of the compacted powder and pre-sintered samples were ~80% and ~78%, respectively. These values concur with the theoretical reduction percentages, wherein $Co_3O_4$, $Fe_2O_3$, and NiO were expected to be fully reduced to metallic Co, Fe, Ni while $Mn_2O_3$ is partially reduced to MnO. The slightly lower reduction percentage of the pre-sintered sample can be attributed to the slower reduction kinetics. This limited kinetics might be due to the small porosity (~2%), and particularly to the limited connected pores that act as a fast diffusion path for both inbound diffusion of $H_2$ and outbound diffusion of water [37]. The HEXRD profiles of the reduced compacted powder and the pre-sintered samples are presented in **Figure 4b**. In both cases the reduced samples consisted of FCC metallic phase and MnO, both with Fm-3m space group (**Table 3**). $Mn_2O_3$ was only reduced to MnO for both sample types during HyDR, due to the relatively high thermodynamic stability of MnO [38]. Notably, a small but distinct fraction of BCC phase (~1 wt.%) was detected in the reduced pre-sintered sample, despite the prevalence of the strongly FCC phase stabilizing elements such as Co and Ni. The existence of BCC phase can be attributed to a depletion of Co and Ni locally, where in the vicinity of Fe-MnO interface. Such a local chemical heterogeneity makes the BCC structure thermodynamically favorable.

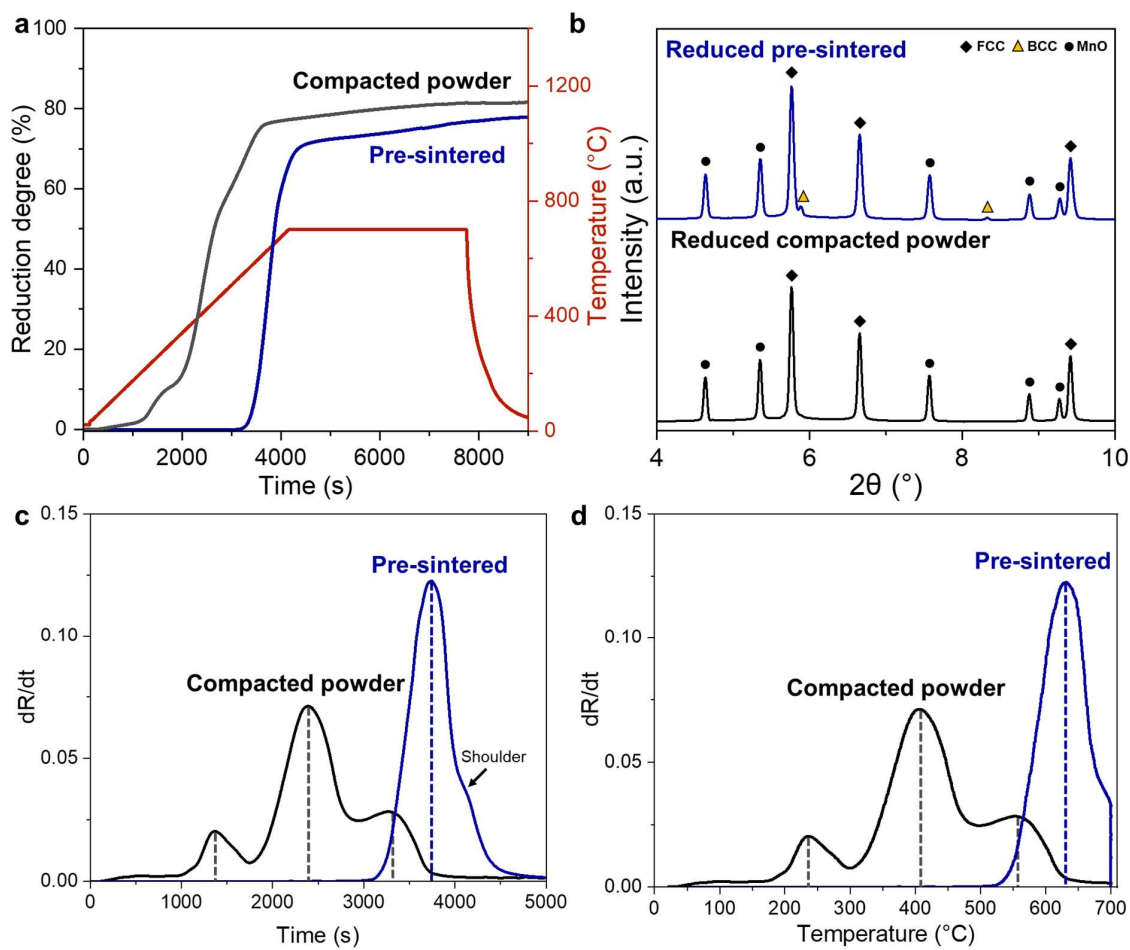

**Figure 4**. (a) Thermogravimetric analysis (TGA) of compacted powder and pre-sintered samples during hydrogen-based direct reduction (HyDR) with a heating rate of 10 °C/min and 1 h holding time at 700 °C. (b) High-energy synchrotron X-ray diffractograms of the HyDR compacted powder and pre-sintered samples reduced at 700 °C. Reduction rate during HyDR of the compacted powder and pre-sintered samples as a function of (c) time and (d) temperature.

**Table 3**. Phase fractions (wt.%) in reduced samples, measured by HEXRD with quantitative phase analysis using Rietveld refinement.

| Sample | FCC | MnO | BCC |
|---|---|---|---|
| Compacted powder | 70±1.0 | 30±0.7 | -- |
| Pre-sintered | 69±1.1 | 30±0.8 | 1±0.2 |

The variation of the reduction rate (the first derivative of the reduction degree) with time and temperature for both, the compacted powder and for the pre-sintered sample is presented in **Figure 4c** and **d**, respectively. The reduction of the compacted powder sample exhibits three distinct peaks, whereas the pre-sintered sample shows only a single dominant peak. Having several peaks in the derivative curves corresponds to reduction rates peaking at different temperatures, depicting different reactions: either a preceding sub-reduction steps, such as $Fe_2O_3$ to $Fe_3O_4$ or a metallization steps, *e.g.*, NiO to Ni. The latter distinction is generally quite relevant in the field of hydrogen-driven reduction. This is because gradual oxygen depletion is, from kinetic, mechanical, and thermodynamic perspectives, profoundly different from discontinuous phase transformation phenomena, that typically occur under the chemically-driven boundary conditions during HyDR.

Interestingly, it can be seen that although two phases (halite and spinel) were formed after sintering, there is only one major peak with a little shoulder towards the final stages of the reduction process (**Figure 4c**). This indicates a concurrent reduction of spinel and halite phases. As a result, the peak reduction rate of the pre-sintered sample (0.12 $s^{-1}$) is ~70% higher than that of the compacted powder (0.07 $s^{-1}$) due to the diffusion-dependent reduction reactions taking place at higher temperatures.

The microstructure of the reduced compacted powder and pre-sintered samples is presented in **Figure 5**. Despite notable differences in the initial porosity between the compacted powder (~26%) and pre-sintered samples (~2%), both exhibited similar porosity (~15%) after HyDR at 700 °C for 1 h, as estimated using ImageJ. These porous microstructures reflect an interplay between pore formation, growth, and coalescence associated with hydrogen reduction and oxygen removal [39]. Additionally, pore shrinkage occurs due to the solid-state sintering that takes place during the reduction process (not to be confused with the sintering prior to reduction which was meant to create a chemically mixed oxide state) [30]. In both samples, a dual-phase microstructure consisting of metallic FCC and MnO phases was observed, agreeing well with the HEXRD results. Annealing twins were observed in the FCC metallic phase, as shown in **Figure 5b**. The EDS maps presented in **Figure 5d**-**h** and **Figure 5l-p** for the reduced compacted powder and pre-sintered samples, respectively, show that Fe, Co, and Ni constitute the solid-solution FCC metallic major phase while the rest is an unreduced MnO minority phase.

A major difference of the two samples lies in the distribution and morphology of MnO phase. The microstructure of compacted powder sample after reduction is more heterogeneous, as there are many large and roughly equiaxed MnO agglomerates (up to several microns in size). In contrast, all the MnO particles is more uniformly spread in the FCC matrix of the pre-sintered sample. This difference likely derives from the microstructure of the initial oxide state (**Figure 3**). The MnO agglomerates resemble relatively large $Mn_2O_3$ particles (~1 μm) in the initial compacted powder. Moreover, the sequential reduction of individual oxides leads to heterogenous porosity. In contrast, the simultaneous reduction of halite and spinel phases in the pre-sintered sample results in a more uniform porosity (**Figure 5**). Furthermore, by observing the reduced pre-sintered sample microstructure at a low magnification, it is possible to discern its morphological similarities to the initial dual-phase microstructure of oxides, in term of the size and distribution of metallic (brighter) and oxide (darker) regions. This pattern is attributable to the fact that only spinel contained Mn-oxide that could only be partially reduced and remain as MnO in the final microstructure. Additionally, the presence of a small fraction of BCC phase – only in the reduced pre-sintered sample – provides further evidence to variations in microstructural evolution between the powder and pre-sintered samples subjected to the same HyDR process.

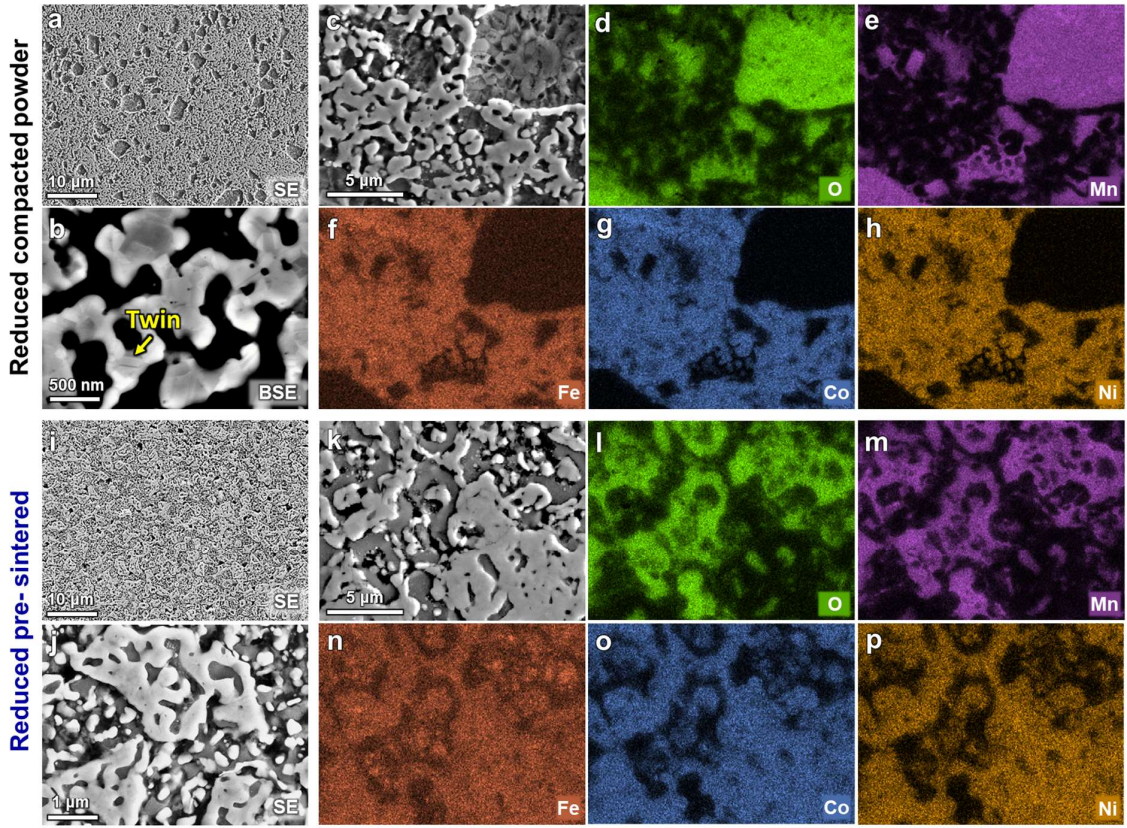

**Figure 5**. Microstructural analysis of samples after hydrogen-based direct reduction (HyDR) at 700 °C at a heating rate of 10 °C/min. Reduced compacted powder samples: SEM micrographs at (a) low and (b) high magnifications. (c) SEM micrograph and corresponding EDS elemental maps of (d) O, (e) Mn, (f) Fe, (g) Co, and (h) Ni. Reduced pre-sintered samples: SEM micrographs at (i) low and (j) high magnifications. (k) SEM micrograph and corresponding EDS elemental maps of (l) O, (m) Mn, (n) Fe, (o) Co, and (p) Ni.

To resolve the small amount of BCC phase observed in the HEXRD analysis, EBSD measurements were performed on the reduced pre-sintered sample, as presented in **Figure 6**. The phase map (**Figure 6a** and **b**) confirmed three distinct phases, namely FCC, BCC, and MnO, in the microstructure. It is evident that the fine Fe (BCC) grains are always adjacent to and typically surrounded by MnO, as revealed in the example marked by the circle in **Figure 6b**. Note that these fine BCC grain may also contain a minor amount (<10%) of Co, and Ni which is insufficient for FCC stabilization. Furthermore, the MnO grains are irregularly shaped, and the BCC Fe grains are submicron in size and equiaxed. In addition, the relatively large and irregular-shaped MnO grains contain many subgrains, as can be seen in the IPF map (**Figure 6**). These facts provide insights into how and why BCC phase formed despite the presence of the Co and Ni in the FCC metallic matrix. Since Fe formed the spinel oxide with Mn during sintering, this compound had to undergo a phase separation process as it partially reduces to metallic Fe. Further, as Fe partitions out of the oxide phase it creates lattice defects, causes deformation, and induces intragranular rotations that form subgrains in the MnO. In turn, the partitioned Fe-oxide readily reduces, and metallic Fe nucleates on the surface of the MnO eventually forming small Fe grains. Ultimately, some of the partitioned Fe (that is mostly encased in MnO) ends up being precluded from mixing with sufficient austenite-stabilizing Ni or Co and remains as α-Fe (**Figure 6d**).

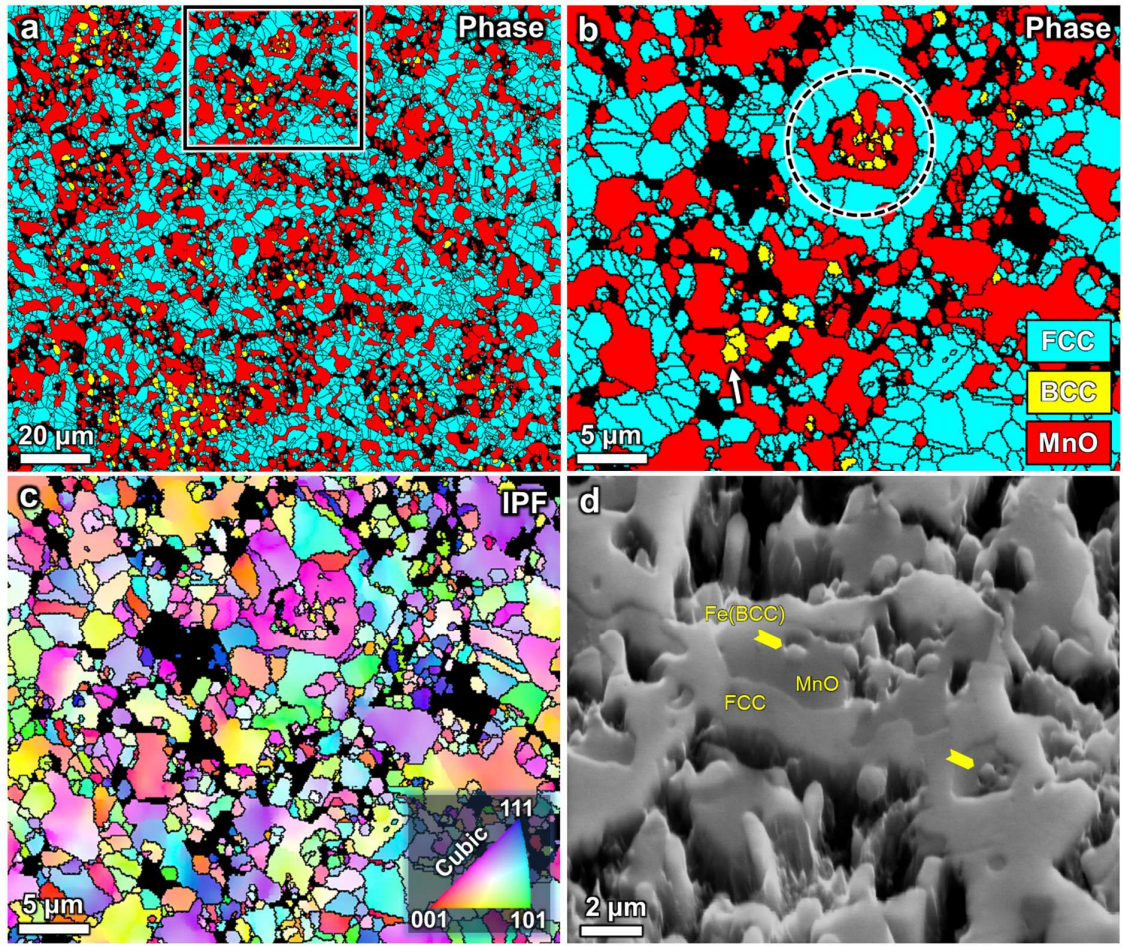

**Figure 6**. EBSD analysis of the reduced pre-sintered sample at 700 °C for 1 h: (a) Phase map showing the FCC, BCC, and MnO phases, (b) higher magnification phase map and corresponding (c) IPF map. White arrow in (b) point to BCC Fe surrounded by MnO. The black lines represent large angle grain boundaries (misorientation >15°). (d) SEM micrograph (70° tilted view) depicting the morphology of fine BCC Fe grains (examples indicated with arrows) surrounded by MnO and metallic FCC phase.

### 3.3 Decoupling thermodynamics and kinetics of reduction of oxide mixtures

To elucidate the thermodynamics of the reduction reactions (*i.e.*, the stability of oxide), it is essential to fully decouple it from kinetic effects. The overall reduction behavior is strongly influenced by sample morphology, particularly the availability of free surfaces for reaction and gas percolation [26,40]. To mitigate the morphological differences between the compacted powder (**Figure 7a**) and the pre-sintered samples, the latter was pulverized by ball-milling for 5 h and 10 h (**Figure 7b** and **c**). Extensive ball-milling for 10 h was required to achieve a particle size distribution (~0.05-5 µm) comparable to that in the original compacted powder sample. XRD analysis of the pulverized pre-sintered samples confirmed that no phase transformations occurred during the milling process (**Figure 7d**). Notably, the pronounced peak broadening observed in the pulverized samples relative to the pre-sintered sample suggests formation of nanoparticles and possible accumulation of lattice strains due to milling-induced crystal defects [41].

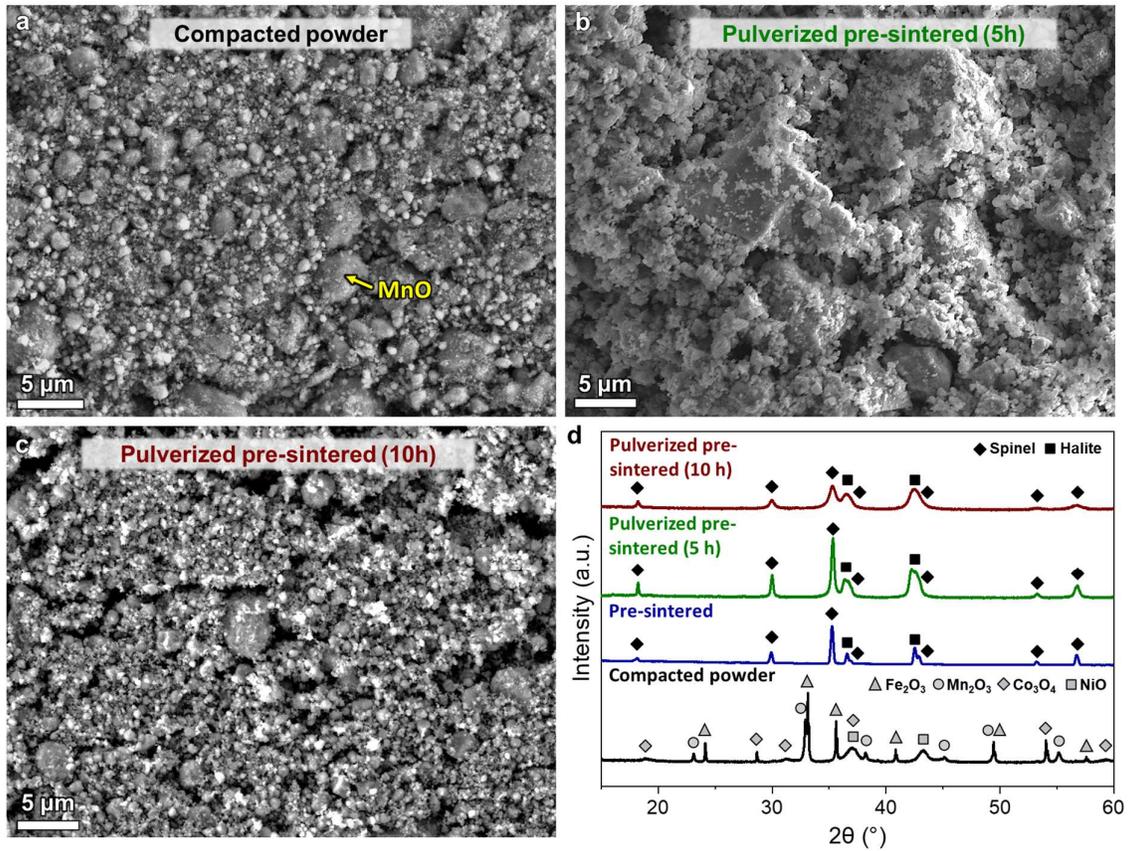

**Figure 7**. SEM micrographs of (a) compacted powder, pulverized pre-sintered sample milled for (b) 5h, and (c) 10h. (d) X-ray diffractogram (λ=0.1.5405 Å) of compacted powder, pre-sintered, pulverized pre-sintered samples.

The TGA results comparing the pulverized pre-sintered samples with the original compacted powder and pre-sintered samples are presented in **Figure 8a**. Pulverizing the pre-sintered sample had a pronounced effect on HyDR, notably on lowering the onset reduction temperatures. Specifically, the onset temperatures decreased from ~525 °C for the pre-sintered sample to ~270 °C and ~175 °C for the pulverized pre-sintered samples after 5 and 10 h of ball-milling, respectively. Despite this shift, the pulverized pre-sintered samples exhibited also a single major kinetic peak (**Figure 8b**), consistent with the XRD results. This suggests that the reduction proceeds through a single dominant mass loss event, similar to the behavior observed for the original pre-sintered sample.

Furthermore, both the peak reduction rate and the temperature at which reduction is completed also decreased following ball-milling: from 630 °C for the original pre-sintered sample to 562 °C and 480 °C after 5 h and 10 h of ball-milling, respectively. These results highlight the significant influence of sample morphology and the initial precursor state on the kinetics and overall reduction behavior of multicomponent oxide mixtures.

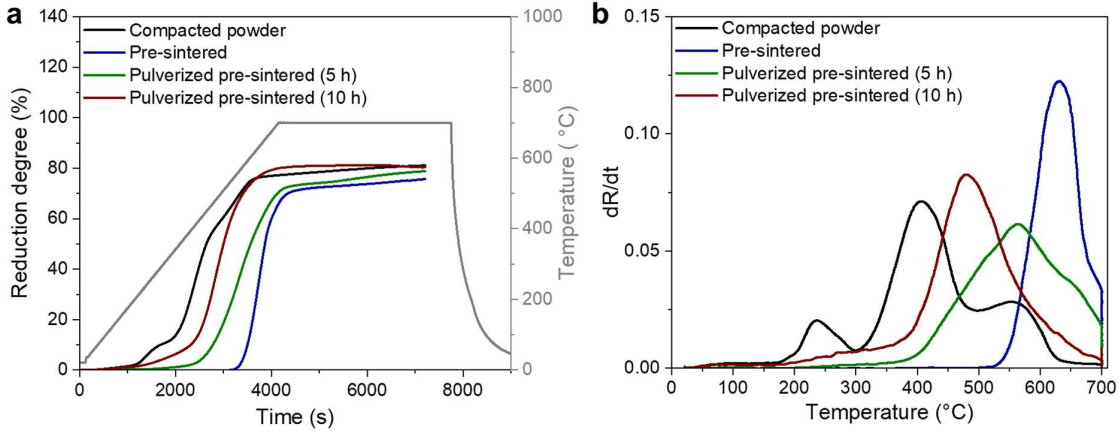

**Figure 8**. TGA reduction tests conducted at 700 °C with a heating rate of 10 °C/min, comparing the compacted powder, pre-sintered, and pulverized pre-sintered samples for 5 h and 10 h: (a) reduction degree (%) as a function of time and (b) reduction rate as a function of temperature.

### 3.4 Correlation between reduction experiments and thermodynamic calculations

Upon sintering the compacted powder, (Co,Ni)O halite and (Fe,Mn)$_3$O$_4$ spinel phases were formed (**Figure 3g**). Thermodynamic calculations reveal that the Gibbs free energies of these mixed halite and spinel phases are notably lower than those of the corresponding Co$_3$O$_4$/NiO and Fe$_2$O$_3$/Mn$_2$O$_3$ oxide precursors, indicating higher thermodynamic stability (**Figure 9a**). The formation of these solid solutions introduces an additional energetic barrier, requiring higher temperatures for their reduction compared to the individual oxides [42]. This thermodynamic insight is consistent with the TGA analysis, where the pre-sintered samples exhibit a delayed reduction onset compared with compacted powder sample (**Figure 8b**).

Furthermore, our previous study demonstrated that the oxygen partial pressure ($p_{O_2}$) serves as an effective predictive metric for assessing the reducibility of the metal oxide mixtures [36]. The phase evolution of the multicomponent oxide mixture as a function of $p_{O_2}$ is depicted in **Figure 9b**. Under relatively oxidizing conditions ($p_{O_2} \approx 10^{-1}$), spinel II (Mn-rich) and halite coexist as stable equilibrium phases. As $p_{O_2}$ decreases, the spinel II becomes thermodynamically unstable as Mn oxide transforms into halite. With further decrease in $p_{O_2}$, the halite phase decomposes into a mixture of spinel I (Fe-rich inverse spinel) and FCC phase.

Eventually, at very low oxygen partial pressure ($p_{O_2} \approx 10^{-32}$), the multicomponent oxide undergoes complete metallization, and the halite phase is fully reduced to an FCC metallic solid solution. This reduction sequence reflects the complex thermodynamic interplay of the phases stability that govern the hydrogen reduction in multicomponent oxide systems.

The elemental partitioning into the FCC metallic phase as a function of $p_{O_2}$ is presented in **Figure 9c**. The plot reveals that Ni and Co oxides are reduced first to form the FCC metallic phase, which is consistent with their lower thermodynamic stability compared with Fe- and Mn-oxides [29]. As $p_{O_2}$ continues to decrease, Fe-oxide is reduced to its metallic state. However, complete metallization of Fe is significantly delayed with the onset of Mn oxide reduction ($p_{O_2} \approx 10^{-23}$). Such a delay suggests a strong interaction between Mn and Fe within the oxide matrix, forming an (Fe,Mn)O halite-type solid solution, which increases the thermodynamic barrier for complete Fe-oxide reduction.

To validate this thermodynamic interpretation, point EDS analysis was conducted on the oxide-rich regions of the reduced compacted powder sample, as shown in **Figure 9d**. The EDS analysis reveals localized Fe enrichment within Mn-rich oxide areas, indicating the presence of a residual FeO phase embedded in a MnO matrix. This finding supports the hypothesis that Fe is partially retained in the oxide phase due to solid solution stabilization by Mn oxide, thereby inhibiting its complete reduction to the FCC metallic phase under the given experimental conditions.

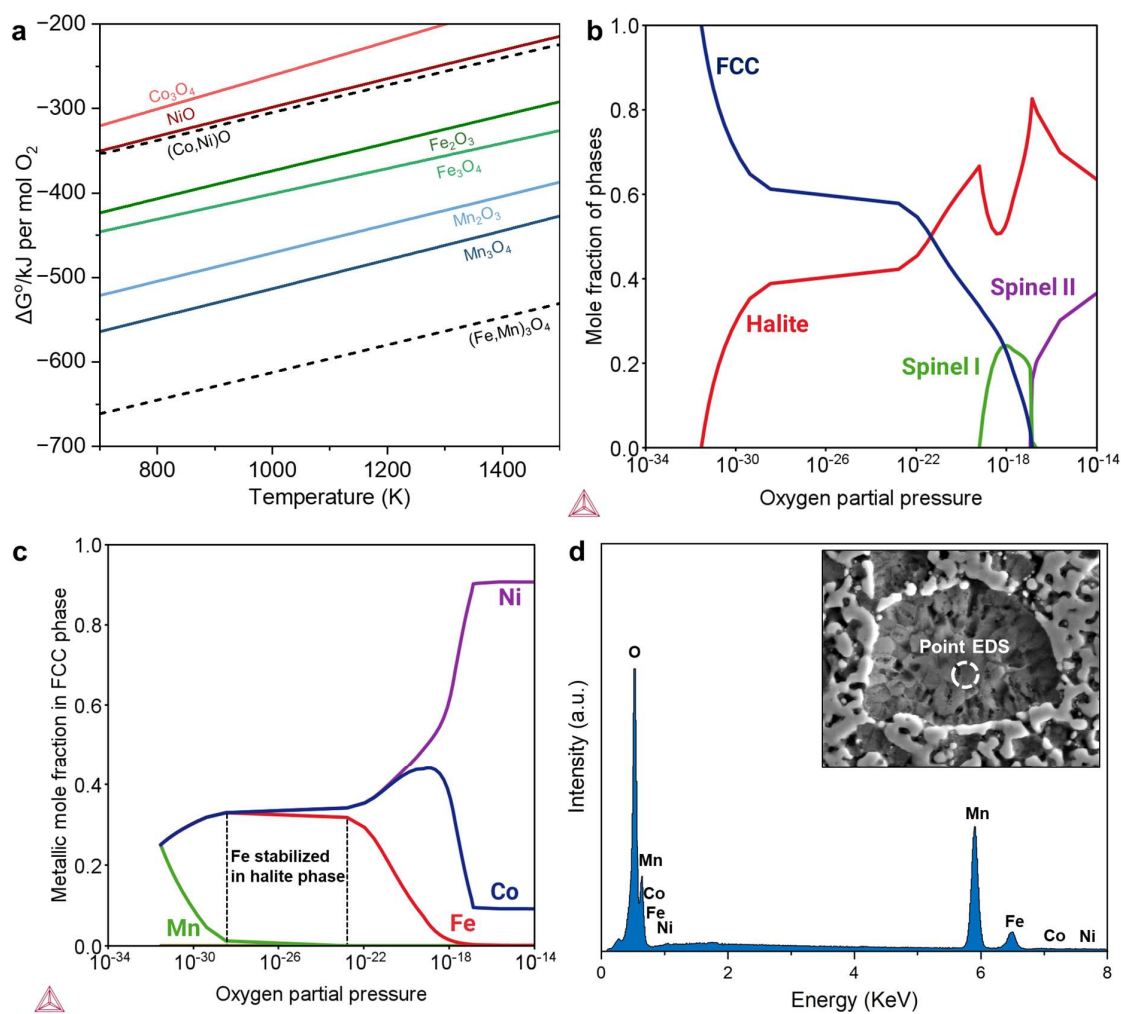

**Figure 9**. (a) Multicomponent-Ellingham diagram for $Co_3O_4$, $Fe_2O_3$, $Mn_2O_3$, NiO, $Fe_3O_4$, $Mn_3O_4$, $(Co,Ni)O$ and $(Fe,Mn)_3O_4$ obtained by thermodynamic calculations using Thermo-Calc. (b) Phase evolution and (c) elemental composition in FCC phase with oxygen partial pressure for a 25Co-25Fe-25Mn-25Ni (at.%) oxide mixture at 700 °C. Spinel I and Spinel II refers to Fe and Mn-rich spinel, respectively. (d) EDS point analysis of an MnO agglomerate in the reduced pre-sintered sample at 700 °C in the oxide region, confirming the presence of residual Fe oxide.

## 4  Conclusions

We investigated the direct reduction behavior using two types of precursor oxide mixtures, namely mechanically mixed compacted powder blends and atomically mixed pre-sintered multicomponent metal oxides. Both types of oxide precursors were derived from $Co_3O_4$, $Fe_2O_3$, $Mn_2O_3$, and NiO, targeting an equiatomic concentration (25 at.%) of the final multicomponent alloy. Despite achieving a similar reduction degree of ~80% after hydrogen-based direct reduction at 700 °C for 1 h, the two samples exhibited distinctly different reduction pathways and microstructures.

The compacted powder precursor underwent sequential reduction of the individual oxides, reflecting the stepwise transformation of these individual constituents. In contrast, the pre-sintered sample, consisting of solid solution oxide phases such as halite and spinel, exhibited a single-step reduction with a high onset temperature (~525 °C). In both cases, the reduction resulted in a two-phase microstructure comprising an FCC metallic phase (70 wt.%) and an unreduced oxide phase (30 wt.%). The FCC metallic phase was a solid solution of Fe, Co, and Ni, while the unreduced oxide was identified as MnO, in accord with thermodynamic calculations. Moreover, the pre-sintered sample contained a minor fraction (1 wt.%) of residual BCC metallic phase. Moreover, pulverizing pre-sintered sample significantly lowered the onset reduction temperatures to 175 °C, attributed to the increased surface area and the disruption of dense, non-porous pre-sintered morphology.

This study showcases a transformative approach to alloy synthesis, enabling one-step direct reduction of multicomponent oxides into compositionally complex materials. By tuning the oxide precursor design and thermodynamic landscape of the oxides, it is possible to simultaneously control reduction pathways, elemental partitioning, and microstructure evolution all within a single process step. This work marks a step towards a novel class of solid-state metallurgical processes capable of producing compositionally complex materials in a single step, paving the way for tailorable alloy manufacturing.

## 5    CRediT authorship contribution statement

**S. S.** Conceptualization, Methodology, Investigation, Formal analysis, Validation, Visualization, Writing – original draft; **B. R.** Conceptualization, Visualization, Supervision, Writing – review & editing; **Y. M.** Conceptualization, Methodology, Supervision, Funding acquisition, Writing – review & editing, **D. R.** Supervision, Funding acquisition, Resources, Writing – review & editing.

## 6    Declaration of competing interest

The authors declare that they have no known competing financial interest or personal relationships that could have appeared to influence the work reported in this paper.

## 7  Acknowledgments

S.S. acknowledges the financial support from Horizon Europe project HAlMan co-funded by the European Union grant agreement (ID 101091936). B.R. is grateful for the financial support of a Minerva Stiftung Fellowship and Alexander von Humboldt Fellowship (Hosted by D.R.). Y.M. acknowledges financial support from Horizon Europe project HAlMan co-funded by the European Union grant agreement (ID 101091936) and the Walter Benjamin Programme of the Deutsche Forschungsgemeinschaft (Project No. 468209039). D.R. acknowledges the financial support from the European Union through the ERC Advanced grant ROC (Grant Agreement No. 101054368). Views and opinions expressed are however those of the author(s) only and do not necessarily reflect those of the European Union or the ERC. Neither the European Union nor the granting authority can be held responsible for them. We acknowledge Benjamin Breitbach for performing the XRD measurements; Jürgen Wichert for his assistance with sample pre-sintering; Rebecca Renz for scientific illustrations (Figure 1).

## 8  Supplementary materials

Supplementary information is available for this paper.